\documentclass{IEEEmce}

\usepackage[colorlinks,urlcolor=blue,linkcolor=blue,citecolor=blue]{hyperref}

\usepackage{upmath}
\usepackage{soul} 
\usepackage[table]{xcolor}
\usepackage{multirow}
\usepackage{hhline}

\jvol{XX}
\jnum{XX}
\paper{8}
\jmonth{xxx/xxx}
\publisheddate{00 xxxx 0000}
\currentdate{00 xxxx 0000}
\jname{IEEE Consumer Electronics Magazine}
\pubyear{2022}
\doiinfo{MCE.2022.Doi Number}

\begin{document}

\sptitle{Department: Head}
\editor{Editor: Name, xxxx@email}

\title{Vulnerabilities and Attacks on CAN-Based 3D Printing/Additive Manufacturing}

\author{Tyler Cultice }
\affil{University of Tennessee, USA}

\author{Himanshu Thapliyal\textsuperscript{1}}
\affil{University of Tennessee, USA}
\markboth{Department Head}{Paper title}

\begin{abstract}
Recent advancements in 3D-printing/additive manufacturing has brought forth a new interest in the use of Controller Area Network (CAN) for multi-module, plug-and-play bus support for their embedded systems. CAN systems provide a variety of benefits that can outweigh typical conventional wire-loom protocols in many categories. However, implementation of CAN also brings forth vulnerabilities provided by its spoofable, destination-encoded shared communication bus. These vulnerabilities result in undetectable fault injection, packet manipulation, unauthorized packet logging/sniffing, and more. They also provide attackers the capability to manipulate all sensor information, commands, and create unsafe operating conditions using only a single compromised node on the CAN network (bypassing all root-of-trust in the modules). Thus, malicious hardware requires only a connection to the bus for access to all traffic. In this paper, we discuss the effects of repurposed CAN-based attacks capable of manipulating sensor data, overriding systems, and injecting dangerous commands on the Controller Area Network using various entry methods. As a case study, we also showed a spoofing attack on critical data modules within a commercial 3D printer.
\end{abstract}

\maketitle
\footnotetext[1]{“This study was prepared under contract with Himanshu Thapliyal, with financial support from the Office of Local Defense Community Cooperation, Department of Defense. The content reflects the views of Himanshu Thapliyal and does not necessarily reflect the views of the Office of Local Defense Community Cooperation.”}
\enlargethispage{10pt}

\chapterinitial{Additive Manufacturing (AM)} is the process of developing 3D objects through an additive, layer-by-layer approach \cite{gibson2021additive}. This process, also known as 3D printing, provides fast prototyping of complex parts and can utilize food, polymers, metals, and living tissue or matter \cite{NGO2018172}. As the complexity and efficiency of 3D printing improves, additional features (including third-party specialized parts) are required for higher integration and more efficient automation. These additional components and communication methods require a robust and strong communication path that minimizes cost of AM and 3D-printing devices.
Controller Area Network (CAN) has recently been increasing in popularity in 3D-printers and other embedded system-based AM machines. Originally designed for vehicular electronic systems, the CAN bus has also found use in other industries like aerospace \cite{Aerospace2006} and robotics \cite{PLATORobotics}. Unlike traditional communication protocols, CAN provides fast, low-error transmission on a two-wire differential bus shared among all devices, allowing easy modularity, providing strong EMI protection, and significantly reducing wire cost. Thus, many printers have begun utilizing this alternative to the traditional protocols to optimize the cost and expand-ability of their application.

\begin{figure}[]
\centering
\includegraphics[width=7.25 cm]{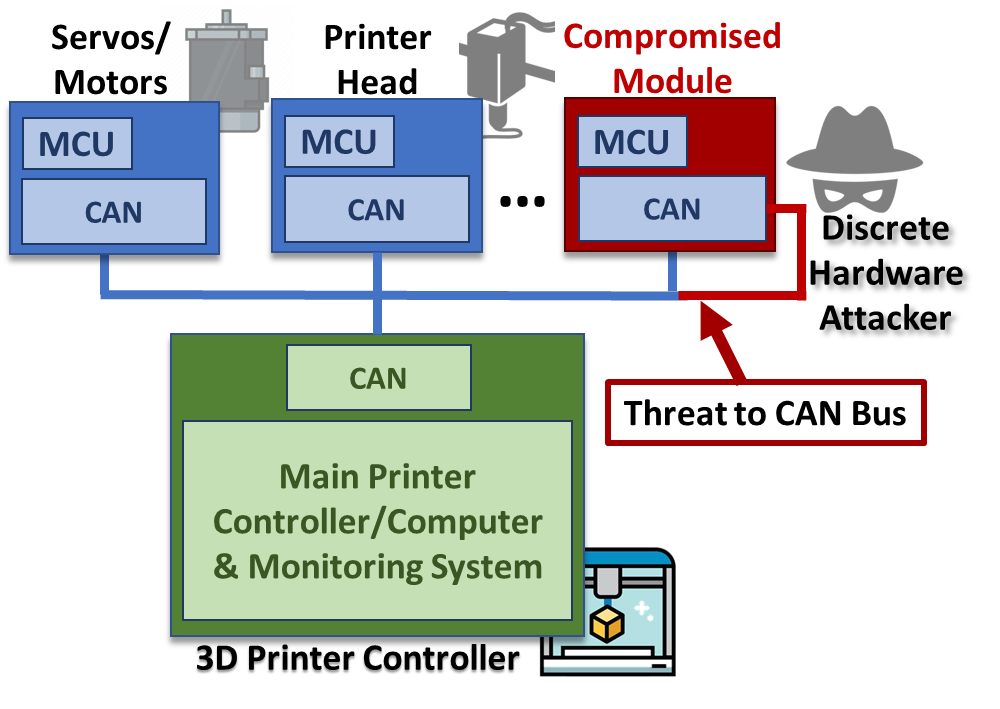}
\caption{Concept of CAN bus hacking through a compromised node, either via hacking an existing node or installing a hardware trojan.}
\label{IntroFig}
\end{figure}

However, CAN's shared, unencrypted communication bus has previously led to many proposed attacks targeting vehicular electronic systems. While vehicles are the primary target of CAN-related attacks, these huge vulnerabilities are also extremely exploitable within CAN AM systems. CAN's destination based IDs mean that any controller can use another connection's ID pattern to easily transmit and receive untraceable messages without any authenticity needed. This means that one compromised node on the CAN network in a 3D printer or printer farm could compromise the entire system's communication.

Thus, discussion of these attacks in context to 3D printing is important to fully understand and counter vulnerabilities as CAN usage continues to grow in AM. An understanding of the capabilities of hardware-related attacks will further promote the use of CAN security frameworks in professional environments, including commercial printers and AM farms. Unlike vehicular CAN systems, the modules being connected to a 3D printer change regularly, thus greatly increasing the exposure and surface of attack from adversaries.

Our contribution in this work is to discuss and demonstrate the capabilities of CAN attacks on additive manufacturing systems that utilize CAN networks for communication between parts/modules. We also evaluate the vulnerabilities and attack surface of a commercial 3D printer that uses a CAN network. Additionally, this article will describe a generalized attack structure that an attacker/eavesdropper would use to perform these attacks discretely through means of compromising a node or new component with malicious hardware, as shown in Figure \ref{IntroFig}. By utilizing a module's physical connection to the bus, attackers can control and manipulate information from all other modules. Only one module vulnerability must be exploited to take control of all bus communication. With this bus-based connection, any information can be sniffed or spoofed from servos, motors, printer head sensors (such as thermistors), controllers, limit switches, and more. Thus, by compromising a module or installing seemingly benign modules containing malicious hardware in unprotected/insecure CAN systems, the many attacks described in this work can be performed against CAN-based 3D printers. The structure of the paper is as follows: Section II provides a background on CAN attacks, Section III describes the attack structure and procedure to identify a point to attack, Section III also provides a demonstration of CAN attacks on a popular commercial printer. Finally, Section V provides concluding thoughts and discussion of future work in AM-based CAN security frameworks.

\section{Background and Related Attacks}
The CAN bus is a multiplexed, message-based communication protocol focused on minimizing wiring costs, providing multi-destination capabilities, and implementing built-in message robustness. These benefits have begun to take the interest of additive manufacturing systems and other cyberphysical systems, especially devices focusing around error-sensitive data transmission with low wire costs between communicating parts. Thus, this protocol has been used in some AM platforms/devices \cite{CANPrintExample} and printing farms \cite{CANFarm} for modularity and simple bus data transfer. Because of the bus-like nature, only a single connection is needed from a micro-controller to have access to any device. Additionally, as a subset of AM, 3D printers utilize the material extrusion process to develop products through an additive process. These systems consist of multiple controllers working in parallel to move parts and extrude filament based on commands transmitted from a central controller. Thus, multiple connection paths must be established between all components to control other modules and provide context (or data) to one another.

\begin{figure}[]
\centering
\includegraphics[width=7.25 cm]{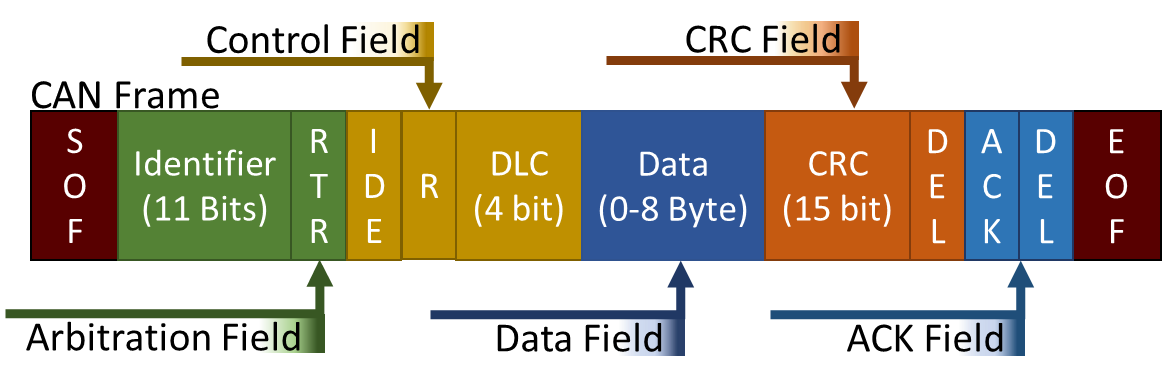}
\caption{Controller Area Network, or CAN, standard message structure.}
\label{CANFrame}
\end{figure}

A CAN "node" consists of a CAN transceiver for physical layer transmission and a driving micro-controller/computer (with CAN controller) for message handling and control. These nodes transmit and receive on the same bus, using IDs contained within the message payload to select recipient(s). This ID can be seen in the arbitration field of Figure \ref{CANFrame}. All CAN devices continually listen/transmit to the bus at all times, but only messages with certain destination IDs (usually bit-masked) will be received and stored by the interested recipients. Thus, given a matching transmission ID, any device on the bus may transmit with any ID regardless of authenticity. Additionally, in the event of message collision, the lower ID message is given transmission priority and retransmission will occur when the bus is free. Nodes may accidentally or purposefully fuzz/denial-of-service (DoS) communication between all devices by overloading the bus or altering the bits/voltage of the bus. Errors generated by faults are mitigated with multiple error correction and detection techniques, including a Cyclic Redundancy Check (CRC) field shown in Figure \ref{CANFrame}.

As described in a technical report by Hartzel et al. \cite{CANAttack}, there are primarily three categories of CAN message assets: (1) private data, (2) safety critical data, and (3) device security data. Private data includes private information of the system's owner that could violate privacy or security, namely printer status information, usage habits, the model being printed, and more. Safety critical data includes data critical to the operation of the system, such as movement or temperature data. Lastly, security data is data used to control security features of the device including 3D printer limit switches, safety switches, air purifiers, etc.

Despite this incredibly sensitive information being transmitted on the bus, there are many ongoing vulnerabilities with CAN: its multicast messaging, lack of message authentication, lack of addressing, common point of entry, lack of encryption, multi-system integration, and limited bandwidth. These vulnerabilities are described in context to additive manufacturing in Table \ref{featureOverview}. Security holes provide access to the insecure CAN bus from an external source, such as wireless or debug ports \cite{CANAtkSurvey}. In 3D printers, these security holes come from unregulated third-party modules, Wi-Fi based monitoring software, network connectivity to the central controller, and other modules that may have outside access.

There are also general categories of threats that attackers may use against CAN as described by \cite{CANSurvey}. Eavesdropping attacks focus on stealing data and analyzing the bus for future attacks, and Data insertion/spoofing attacks focus on generating unauthorized CAN frames in the network by spoofing the destination ID and payload structure of an established CAN communication. Lastly, DoS attacks prevent any particular node (or whole network) from providing service to other nodes. While physical manipulation and message spam are typical DoS methods for CAN, Mukherjee et al. \cite{CANDoS} also demonstrated manipulation of special signals to deny service. Hartzel et al. Finally, \cite{CANAttack} described three additional possible attacker profiles: a thief injecting software to steal data or alter control systems, a manufacturer injecting harmful data/firmware into the bus affecting safe operation, and the system owner accidentally transmitting incorrect data or unsafe devices. Some existing security frameworks designed for vehicle environments have been proposed to combat these in automotive CAN networks \cite{EasySec,CANPQC,CANOldPaper}.

\begin{table}[]
\renewcommand{\arraystretch}{1.4}
\setlength{\tabcolsep}{4pt}
\caption{Attacks and risks on cyberphysical 3D printer based on CAN vulnerabilities in Hartzel et al. \cite{CANAttack}.}
\label{featureOverview}
\centering
\begin{tabular}{|p{3cm}|p{3.55cm}|}
\rowcolor[HTML]{2F5597}
\textcolor{white}{Vulnerability}&\textcolor{white}{Capabilities on CAN Bus}\\
\hline
\rowcolor[HTML]{DAE3F3}
Multi-cast Messaging&Undetectable, passive listening to all messages can occur regardless of recipient(s).\\
\hline
\rowcolor[HTML]{DAE3F3}
No Message Source Authentication/Addressing&Spoofed messages can come from any device or source.\\
\hline
\rowcolor[HTML]{DAE3F3}
Common Point of Entry&Adversary can gain access to all nodes from a single exploit.\\
\hline
\rowcolor[HTML]{DAE3F3}
Lack of Encryption/Data Isolation&Information is not private, can be easily read and fabricated by any device on the bus.\\
\hline
\rowcolor[HTML]{DAE3F3}
Multi-system Integration&Various different systems integrate onto the CAN bus, extending the attack surface.\\
\hline
\rowcolor[HTML]{DAE3F3}
Limited Bandwidth&Easy to denial-of-service safety and security modules.\\
\hline
\end{tabular}
\end{table}

\begin{figure*}[]
\centering
\includegraphics[width=14.5 cm]{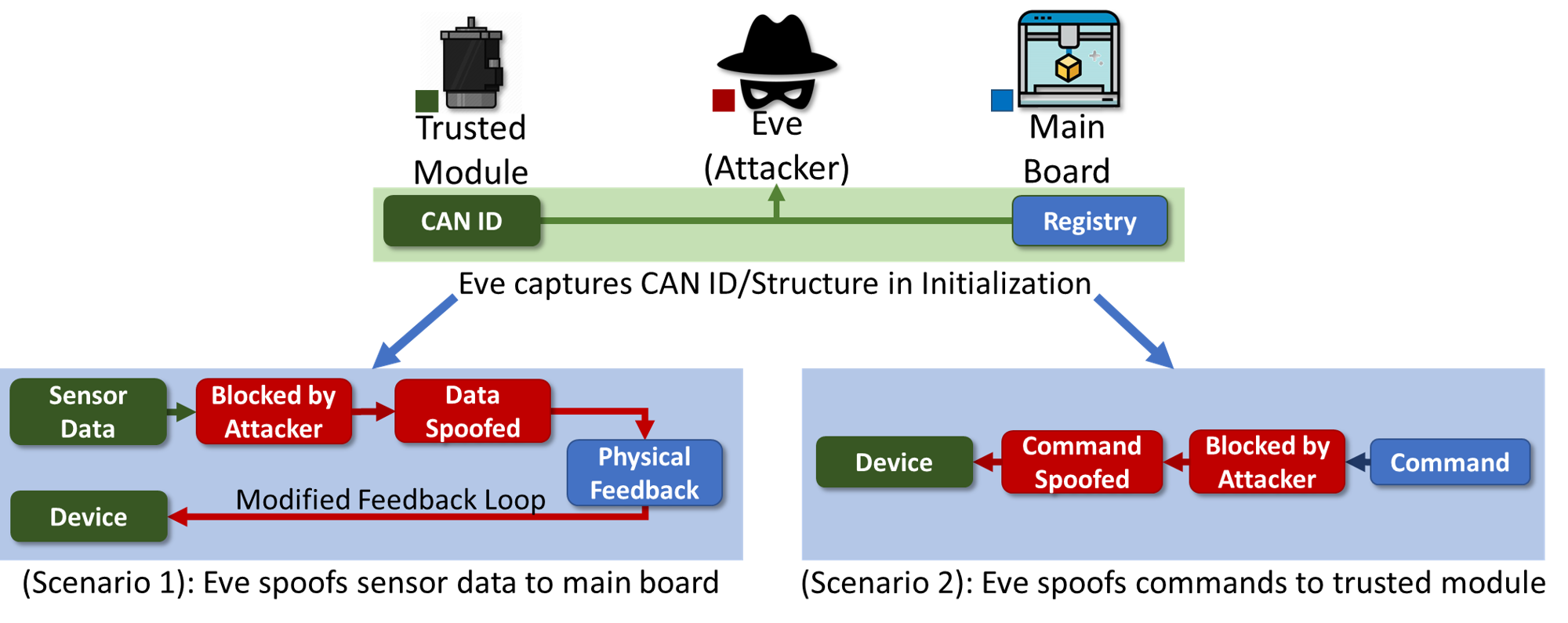}
\caption{Design and structure of a generalized CAN attack on 3D printers. CAN information is captured by the attacker during printer initialization of the modules via CAN. (Scenario 1) Commands from the main board are blocked and spoofed by the adversary to control the modules directly. (Scenario 2) Spoofed sensor data is generated and sent to the main board, causing a change in the response/feedback loop.}
\label{AttackStruct}
\end{figure*}

\section{Attack Structure and Demonstration}
Similarly to vehicular CAN attacks, the structure of 3D printing CAN attacks presents itself in the form of compromising an existing node or including malicious hardware within a new/replacement node. As CAN 3D printing provides bus-based modularity, the information needed to understand the system can be easily captured and reverse engineered during the device initialization and registration period. This is shown in Figure \ref{AttackStruct}. The attack begins with the attacker building context on each device's capabilities and designated CAN bus transmission ID. This will provide data and context on the network that allows for the attacker to know what ID and message types map to which functionalities. The attacker can then manipulate any information through the compromised node after initialization to control or block/DoS any device. This allows attackers to manipulate target/current temperatures, device states, external peripherals (i.e., cameras), positions and motor statuses (x,y,z), limit switches, PID values, and other functionalities that require constant and robust data transmission. A combination of data manipulation and DoS attacks can result in discrete and undetectable damage to the device, such as sending a malicious target temperature to the printer head and spoofing a normal response temperature back to the monitoring software. Thus, with all this information being transmitted through CAN as the primary serial-communication tool, simple attacks like packet injection or replay attacks can easily wrest control of various kinds of CAN data. The data vulnerable in CAN 3D printing devices to simple attack alterations following this structure are given in Table \ref{TableAttacks}. The table is sorted by category of CAN data as per described earlier in section II \cite{CANAttack}, ranging from user's private data (STL files, settings, user accounts, etc.) to even the security data used to monitor other types of attacks (i.e., security cameras, security updates within firmware, etc.).

\begin{table*}[]
\renewcommand{\arraystretch}{1.4}
\setlength{\tabcolsep}{4pt}
\caption{CAN-based 3D-printer data vulnerable to CAN attacks based on category of critical data/functionality.}
\label{TableAttacks}
\centering
\begin{tabular}{|p{2.5cm}|p{3cm}|p{8.5cm}|}
\hline
\rowcolor[HTML]{2F5597}
\textcolor{white}{Category} & \textcolor{white}{Type of Data}&\textcolor{white}{Proposed Method of Attack}\\
\hline
\rowcolor[HTML]{DAE3F3}
& STL/GCode Data&Eavesdropping on positional data sent to motor or printer head modules via CAN bus.\\
\hhline{|>{\arrayrulecolor[HTML]{DAE3F3}}->{\arrayrulecolor{black}}-|-|}
\rowcolor[HTML]{DAE3F3}
&Printer Status & Eavesdropping/Fabricating status information transmitted to controlling or monitoring system.\\
\hhline{|>{\arrayrulecolor[HTML]{DAE3F3}}->{\arrayrulecolor{black}}-|-|}
\rowcolor[HTML]{DAE3F3}
&User Preference/Settings & Eavesdropping on transmitted user settings, such as Wi-Fi networks, account data, or device identifiable information that may be sent via CAN.\\
\hhline{|>{\arrayrulecolor[HTML]{DAE3F3}}->{\arrayrulecolor{black}}-|-|}
\rowcolor[HTML]{DAE3F3}
\multirow{-5.23}{*}{Private Data}&Device Calibration Data & Spoofing and DoS during calibration can result in miscalibrated and out-of-sync modules.\\
\hline
\rowcolor[HTML]{DAE3F3}
& Temperature/Sensor Data&Manipulating reported sensor or temperature readings sent to central controller.\\
\hhline{|>{\arrayrulecolor[HTML]{DAE3F3}}->{\arrayrulecolor{black}}-|-|}
\rowcolor[HTML]{DAE3F3}
&Target Temperature & Spoofing temperature target commands sent to heating modules (i.e., printer head).\\
\hhline{|>{\arrayrulecolor[HTML]{DAE3F3}}->{\arrayrulecolor{black}}-|-|}
\rowcolor[HTML]{DAE3F3}
&Limit Switches/Position & Fuzzing/Spoofing limit and positional data to misalign, miscalibrate, or control device manually.\\
\hhline{|>{\arrayrulecolor[HTML]{DAE3F3}}->{\arrayrulecolor{black}}-|-|}
\rowcolor[HTML]{DAE3F3}
&Safety Device Data & Spoofing/blocking safety features (i.e., air filters, kill switches, fire alarms)\\
\hhline{|>{\arrayrulecolor[HTML]{DAE3F3}}->{\arrayrulecolor{black}}-|-|}
\rowcolor[HTML]{DAE3F3}
\multirow{-7}{*}{Safety Critical Data}&General Data/Commands & Overwhelming/Manipulating the physical bus via DoS to block vital commands or data.\\
\hline
\rowcolor[HTML]{DAE3F3}
& Cameras/Monitoring & Spoofing or DoS attacks on printer farm monitoring devices, such as cameras or sensors in \cite{CANFarm}.\\
\hhline{|>{\arrayrulecolor[HTML]{DAE3F3}}->{\arrayrulecolor{black}}-|-|}
\rowcolor[HTML]{DAE3F3}
& Module Firmware & Software updates allow firmware flashing through CAN \cite{CANFlash}.\\
\hhline{|>{\arrayrulecolor[HTML]{DAE3F3}}->{\arrayrulecolor{black}}-|-|}
\rowcolor[HTML]{DAE3F3}
\multirow{-3.5}{*}{Security Data}&Settings/Config Data&Spoofing/Sniffing device data (i.e., device settings, security data).\\
\hline
\end{tabular}
\end{table*}

\begin{figure}[h]
\centering
\includegraphics[width=7.25cm]{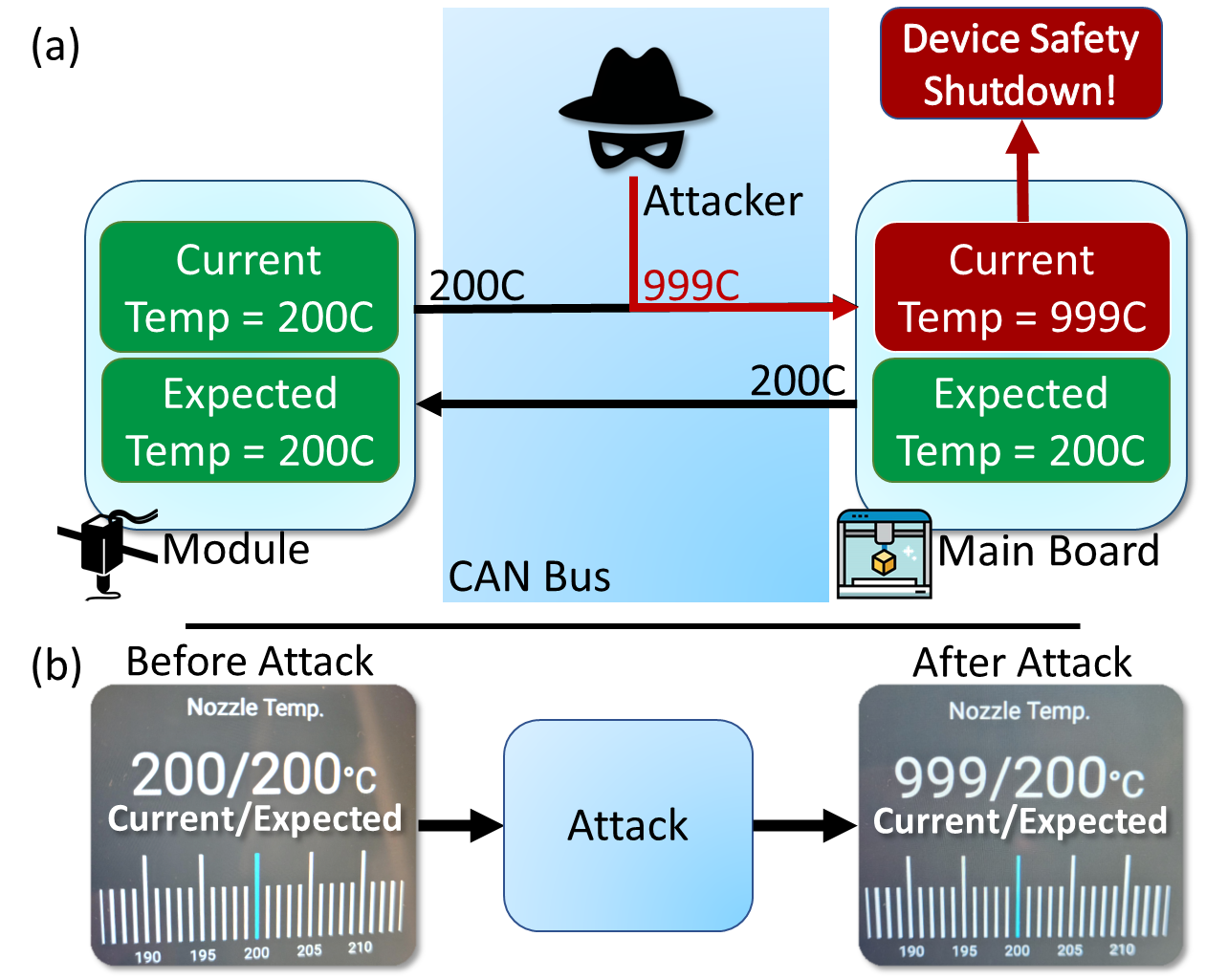}
\caption{(a) Graphic of a temperature spoofing attack on the CAN connection between the main board and a genuine nozzle thermistor sensor. After transmitting a spoofed 999 Celsius response, the system reacts to a seemingly unsafe condition despite the nozzle operating normally. (b) Graphical output of the received module temperature by the central controller/main board is shown changing from the normal 200C to 999C, locking up the entire system.}
\label{FigInject}
\end{figure}

\subsection{Attack Demonstration}
To demonstrate the impact of these CAN attacks, we demonstrated various attacks on a commercial 3D printer seen in various AM-based projects \cite{PrinterExample,PrinterExample2}. This printer uses CAN to transmit data and commands to each of the modules in the system (i.e., printer head, limit switches, etc.). While this printer utilizes main-board control of some functionalities instead of CAN, many attacks can still be performed by spoofing, blocking, or manipulating any of the transmitted data sent over CAN. Our attack setup consists of a SAM C21 micro-controller to probe an empty port on the CAN Hub, an easily accessible component of the 3D printer meant for plug-and-play capabilities. This micro-controller plays the role of a malicious node in the system, which could be interpreted as either a hardware trojan (plugged into the CAN hub) or an existing hacked module. We then overloaded the bus with "current temperature" data packets containing the temperature 999C, despite the sensor's actual reading of 200C. This caused the printer to instantly fault and shut down, as that temperature was well out of the range of normal operation. The concept of the attack and resulting printer interface (after changing the temperature) can be seen in Figure \ref{FigInject}. Using this same attack scheme, we were also capable of performing many of the other proposed CAN 3D-printer vulnerabilities in Table \ref{TableAttacks}, such as miscalibrating the system, spoofing limit switches, transmitting dangerous target conditions to modules (while disabling safety features meant to disable the system in dangerous conditions), overwriting module software (through their reflashing through CAN system), and maliciously controlling/disabling the air filter/enclosure. Our setup was capable of manipulating the various feedback loops of movement, temperature, and calibration to both cause and hide unsafe conditions for the printer and the user of the printer. These attacks can be combined to further manipulate the printer, monitoring software, or other safety/security features included in the device. For example, our attacks were capable of denying the central controller the ability to turn on the nozzle's fan while setting the target temperature to dangerously high values. Additionally, we were able to set the nozzle to temperatures capable of melting the filament while shutting off the air purifier.

Additionally, these attacks are not specific to any particular 3D printer either. Without proper CAN security, any AM device utilizing a CAN bus may be vulnerable to these attacks to lesser or higher amounts. With the same micro-controller and attack structure, our attack may be utilized on another AM device with ease. The attack surface to perform our work on continues to grow due to increasing amounts of IoT capabilities, frequent part swapping, and external connections to CAN-enabled monitoring devices or printer modules. As a result, the severity of CAN related attacks on increasingly internet-connected cyberphysical AM devices also continues to increase.
\section{Concluding Thoughts}
In this article, we demonstrate that attacks on 3D printing CAN networks can perform severely malicious effects on commercial and personal additive manufacturing systems by manipulating the transmitted data between the main board and modules. Our work determined that various subsystems of the 3D printer were vulnerable to CAN attacks, including privacy and safety information. These shared communication paths provide attackers with access to most of the modules and functionalities, or even other printers and monitoring devices in a 3D printing farm. Due to inheritance of the vulnerabilities associated with CAN, an adversary can easily compromise an entire 3D printing device or farm through a single exploit to risk private, safety critical (including the user themselves), and device security data.

In future work, Security features should be explored to mitigate the threat of adversaries without significantly reducing the benefits of CAN in the printer. Despite many vehicular CAN security frameworks have been proposed in the past, additive manufacturing modules provide a new challenge to CAN security due to its focus on easy part swapping, modularity, and plug-and-play. Most existing CAN security concepts result in problems which significantly diminish the benefits that CAN brings to the additive manufacturing market. However, by focusing on and proposing future AM security, even modular/plug-and-play printers can provide sufficient privacy and security between the modules without losing the benefits of CAN.

While our attacks demonstrate strong risk to safety and privacy, this does not conclude the removal of CAN from our AM and 3D printing systems. To fully secure The future of CAN in AM, it will require manufacturers and researchers to revisit CAN security and attacks in the context of modular, dynamic 3D printers CAN networks. Thus, CAN will require development into a strong design around security, reliability, and privacy for both commercial farms and personal printers alike.



\bibliographystyle{IEEEtran}
\bibliography{references}

\begin{IEEEbiography}{Tyler Cultice}{\,} is currently working toward the Ph.D. degree in electrical and computer engineering with the University of Tennessee, Knoxville, TN 37996, USA.  Contact him at tcultice@vols.utk.edu.
\end{IEEEbiography}

\begin{IEEEbiography}{Himanshu Thapliyal}{\,} is an Associate Professor with the Department of Electrical Engineering \& Computer Science, University of Tennessee, Knoxville, TN 37996, USA. Contact him at hthapliyal@utk.edu.
\end{IEEEbiography}

\end{document}